\documentclass[12pt,preprint]{aastex}

\shorttitle{Close binary stars.~XI}
\shortauthors{Pribulla \& et al.}

\begin{document}

\title{Radial Velocity Studies of Close Binary
Stars.~XI\footnote{Based on the data obtained at the David Dunlap
Observatory, University of Toronto.}}

\author{Theodor Pribulla}
\affil{Astronomical Institute, Slovak Academy of Sciences \\
059 60 Tatransk\'a Lomnica, Slovakia}
\email{pribulla@ta3.sk}

\author{Slavek M. Rucinski,
        Wenxian Lu\altaffilmark{2},
        Stefan W. Mochnacki,\\
        George Conidis,
        R.M. Blake\altaffilmark{3},
        Heide DeBond, J.R. Thomson}
\affil{David Dunlap Observatory, University of Toronto \\
P.O.~Box 360, Richmond Hill, Ontario, Canada L4C~4Y6}
\email{(rucinski,mochnacki,conidis,debond,jthomson)@astro.utoronto.ca}

\author{Wojtek Pych}
\affil{Copernicus Astronomical Center, Bartycka 18, 00--716 Warszawa, Poland}
\email{pych@camk.edu.pl}

\author{Waldemar Og{\l}oza}
\affil{Mt. Suhora Observatory of the Pedagogical University\\
ul.~Podchora\.{z}ych 2, 30--084 Cracow, Poland}
\email{ogloza@ap.krakow.pl}

\author{Michal Siwak}
\affil{Astronomical Observatory, Jagiellonian University, ul. Orla 171,
       30--244 Cracow, Poland}
\email{siwak@oa.uj.edu.pl}

\altaffiltext{2}
{Current address: Department of Geography and Meteorology and
 Department of Physics and Astronomy, Valparaiso University, Valparaiso,
 IN 46383, U.S.A. e-mail: Wen.Lu@valpo.edu}
\altaffiltext{3}
{Current address: Pisgah Astronomical Research Institute,
 1 PARI Dr. Rosman, N.C., 28772, U.S.A. e-mail: mblake@mail.pari.edu}

\begin{abstract}
Radial-velocity measurements and sine-curve fits to the orbital radial
velocity variations are presented for ten close binary
systems: DU~Boo, ET~Boo, TX~Cnc, V1073~Cyg, HL~Dra, AK~Her, VW~LMi,
V566~Oph, TV~UMi and AG~Vir. By this contribution, the DDO program has
reached the point of 100 published radial velocity orbits.
The radial velocities have been determined using an improved
fitting technique which uses rotational profiles to approximate
individual peaks in broadening functions.

Three systems, ET~Boo, VW~LMi and TV~UMi, were found to be
quadruple while AG~Vir appears to be a spectroscopic triple.
ET~Boo, a member of a close visual binary with $P_{vis} = 113$ years,
was previously known to be a multiple system, but we show that
the second component is actually a close, non-eclipsing binary.
The new observations enabled us to determine the spectroscopic
orbits of the companion, non-eclipsing pairs in ET~Boo and VW~LMi.
The particularly interesting case is VW~LMi, where the period of
the mutual revolution of the two spectroscopic binaries is only
355 days.

While most of the studied eclipsing pairs are contact binaries,
ET~Boo is composed of two double-lined
detached binaries and HL~Dra is single-lined
detached or semi-detached system. Five systems of this group
were observed spectroscopically before:
TX~Cnc, V1073~Cyg, AK~Her (as a single-lined binary),
V566~Oph, AG~Vir, but our new data are of much higher quality than
the previous studies.
\end{abstract}

\keywords{ stars: close binaries - stars: eclipsing binaries --
stars: variable stars}

\section{INTRODUCTION}
\label{sec1}

This paper is a continuation of a series of papers (Papers
I -- VI and VIII -- X) of
radial-velocity studies of close binary stars
\citep{ddo1,ddo2,ddo3,ddo4,ddo5,ddo6,ddo8,ddo9,ddo10} and presents
data for the tenth group of ten close binary stars observed
at the David Dunlap Observatory. For technical details and
conventions, and for preliminary estimates of uncertainties,
see the interim summary paper \citet[Paper VII]{ddo7}.

In this paper, we make use of broadening functions extracted not
only from the region of the Mg~I triplet at 5184~\AA,
as in previous papers, but
also from two regions containing telluric features centered at
6290~\AA\ and 6400~\AA. These experimental setups were used
because of concerns about flexure effects in our spectrograph.
While this experiment provided a good check on the stability of
our radial-velocity system and -- to a large extent -- alleviated
our concerns, we found that the stellar lines in these two
regions were generally too weak to replace the 5184~\AA\ feature on a
routine basis. The broadening functions (from now on
called BF's) based on the 6290~\AA\ and 6400~\AA\ observations
were rather poorly defined, especially for earlier spectral types;
this was mostly due to the low efficiency of our diffraction
grating in the red region. As the result, the BF's for AG~Vir and
DU~Boo were poor, with the secondary component almost
undetectable. Thus, in the end, we have returned to the 5184~\AA\
region for the subsequent observations. The flexure tests based on
our telluric-lines template (Regulus = HD87901) have shown that the
standard wavelength calibrations provide a reasonable stability of
our radial-velocity system with the largest deviations within
$\pm$3 km~s$^{-1}$. The broadening functions obtained in the red
region were used in the present study only to augment the data for
the quadruple systems, ET~Boo, VW~LMi and TV~UMi, and only
for observations at critical orbital phases of long-period systems
when any spectrum was of use in defining the orbit.

In August 2005, a new grating with 2160 l/mm was acquired to replace
the previously most frequently used grating with 1800 l/mm which after
many years of use lost its original efficiency. Thus, unfortunately, due
to the changes described above, in combination with the necessarily
very extended time coverage for triple and quadruple systems,
the present dataset is the least homogeneous
since the start of this series of studies.
The BF's used here were extracted from spectra obtained with four different
CCD detectors and two different diffraction gratings. This lack of homogeneity
does not seem to affect the final data which have uncertainties
similar to previously reported in this series of investigations.

Selection of the targets in our program remains quasi-random: At a given time,
we observe a few dozen close binary systems with periods usually
shorter than one day, brighter than 10 -- 11 magnitude and with
declinations $>-20^\circ$; we publish the results in groups of ten systems
as soon as reasonable orbital elements are obtained from measurements evenly
distributed in orbital phases.
Whenever possible, we estimate spectral types of the program stars
using our classification spectra obtained with a grating of
600 l/mm over a range of 635 \AA\ or 890 \AA\ (depending
on the CCD detector) centered at 4200 \AA. Our 
classifications are based on comparison with several spectral
standards of the MK system observed on the same nights.
They are compared with the
mean $(B-V)$ color indices usually taken from the Tycho-2
catalog \citep{Tycho2} and the photometric estimates of the spectral types
using the relations published by \citet{Bessell1979}.

The radial velocity (hereafter RV) observations reported in this paper
have been collected between June 1997 and September 2005.
The ranges of dates for individual systems can be
found in Table~\ref{tab1}.
The present group contains 3 quadruple systems,
ET~Boo, VW~LMi and TV~UMi, whose complex nature had been noticed
several years ago, but whose full orbital solutions required extended monitoring.

This paper is structured in a way similar to that of previous
papers, in that most of the data for the observed binaries are in
two tables consisting of the RV measurements in
Table~\ref{tab1} and their preliminary sine-curve solutions
in Table~\ref{tab2}. RVs and corresponding spectroscopic
orbits for all ten systems are shown in Figures~\ref{fig1} to \ref{fig3}.
In this paper we changed the way how RVs
are determined from the broadening
functions: Instead of Gaussians, we now use single or double
rotational profiles to fit the peaks in the broadening functions.
This approach, which is described in Section~\ref{rot},
gives much better results with smaller random errors. We stress that
this is still not a full modeling of the broadening function
shape (as attempted in \citet{ruc92,ahvir,wuma})
which would be an optimal approach, but a convenient and better working
(than Gaussians) tool. The measured RVs are listed in Table~\ref{tab1}
together with weights, determined as $1/\sigma^2$,
as based on individual determinations of central
component velocity. This weighting scheme, which accounts for differences
in relative quality of observations, improves the quality of
the orbital solutions.

\setcounter{footnote}{3}    % confusion of footnote numbers in the title

The data in Table~\ref{tab2}
are organized in the same manner as in previous papers.
In addition to the parameters of spectroscopic orbits,
the table provides information about the relation between
the spectroscopically observed epoch of the primary-eclipse T$_0$
and the recent photometric determinations in the form of the $O-C$
deviations for the number of elapsed periods $E$. For HL~Dra the
reference ephemeris is taken from the Hipparcos Catalogue;
for DU~Boo from \citet{prib2005}; for the rest
of the systems, ephemerides given in the on-line version of
``An Atlas O-C diagrams of eclipsing binary
stars''\footnote{http://www.as.wsp.krakow.pl/ephem/} \citep{Kreiner2004}
were adopted.
Because the on-line ephemerides are frequently
updated, we give those used for the computation of the $O-C$
residuals below Table~\ref{tab2} (status as in February 2006).

The values of $T_0$ refer to the deeper eclipse which for W-type
systems corresponds to the lower conjunction of the more massive
component; in such cases the epoch is a non-integer number.
In the cases of ET~Boo and VW~LMi, where observations
covered several years and photometric data have been rather scanty, we
optimized not only $T_0$, but also the orbital period.

Table~\ref{tab2} contains our new spectral classifications of the program
objects. Independent classification was done for all systems except
TX~Cnc. Section~\ref{sec2} of the paper contains brief summaries of
previous studies for individual systems and comments on the new data.
The novel technique of fitting the rotational
profiles to peaks in the BF's is described in Section~\ref{rot}.
Examples of BF's of individual systems extracted from spectra observed
close to quadratures are shown in Fig.~\ref{fig4}.

Similarly as in our previous
papers dealing with multiple systems, RVs for the eclipsing
pair were obtained from BF's with the additional peaks removed.
This task was performed by first fitting multiple Gaussian profiles to the
combined BF's and then removing the signatures of the third (and
sometimes fourth) component. While the final RVs of
the close pair were determined by rotational profile fitting to
such ``cleaned'' profiles,
the velocities of well separated and slowly rotating components
of the additional components were determined by the Gaussian fits
(Table~\ref{tab3}). Because the BF technique actually produces
Gaussians for intrinsically sharp signatures
with $\sigma \simeq 15$ km~s$^{-1}$, this approach
is internally consistent.

\section{RESULTS FOR INDIVIDUAL SYSTEMS}
\label{sec2}

\subsection{DU~Boo}
\label{duboo}

Photometric variability of DU~Boo was discovered by the
Hipparcos satellite \citep{hip} where the star was classified as
ellipsoidal variable of the A2 spectral type. Later,
\citet{gomgar1997} observed the system photometrically and found
that it is an eclipsing binary with a large O'Connell effect
amounting to the difference
$Max.II - Max.I = 0.10$ mag in the $V$ passband.
It is interesting to note that light-curve asymmetry and the
associated surface inhomogeneities have been very stable since the
time of the Hipparcos discovery; this indicates that solar-type
dark-spots paradigm does not apply in this case.
Recently, \citet{prib2004} analyzed the $UBV$ photometry and found
that DU~Boo is a relatively long-period (1.0559 day)
contact binary showing total eclipses; the derived
photometric mass ratio was found to be $q = 0.194(2)$.
Our spectroscopic mass ratio $q = 0.234(35)$ is
consistent with the photometric determination,
which documents reliability of photometric mass ratios
derived from timing of the inner eclipse contacts for
contact binaries showing total eclipses \citep{MD72a,MD72b}.

The large O'Connell effect is reflected in the extracted broadening
functions of DU~Boo. While the primary component shows undisturbed
broadening functions around quadratures, close in shape
to the theoretical rotational profiles
(Section~\ref{rot}), the BF profile for the
secondary is always very deformed. This causes distortions
of the observed RV curve and adversely affects
the solution for the mass-center velocity. The peaks of the
BF's are not fully separated supporting the photometric
solution of DU~Boo as a contact binary. The system is clearly
of the A-type with the more massive component eclipsed at the
deeper minimum.

By combining our spectroscopic results with
the inclination angle $i = 81.5\degr$ \citep{prib2004}, we obtain
the total mass of the system $M_1+M_2 = 2.56 \pm 0.07 M_\odot$.
Our new spectral type estimate of A7V is definitely later than the spectral
type given in the Hipparcos catalogue (A2). The mean Tycho-2 color
$(B-V) = 0.31$ is in better accord with our determination of the
spectral type and requires only a small interstellar extinction.
The orbital period of the system (1.0559 days) is rather long for
a contact binary of A7V spectral type which indicates that the components
of DU~Boo may be evolved. The Hipparcos parallax
is relatively small $\pi =  2.58 \pm 1.03$ and not precise enough for
determination of the system luminosity.

\subsection{ET~Boo}
\label{etboo}

The photometric variability of the star was discovered
by the Hipparcos satellite \citep{hip}; where it is cataloged
as a $\beta$~Lyrae eclipsing binary of the F8
spectral type. ET~Boo is a known member of the visual pair COU~1760
(ADS~14593+4649), with the orbital period about 113 years and
the magnitude difference of
$\Delta V$ = 0.86 (see Sixth Catalog of Orbits of Visual Binary Stars,
currently available only in the electronic form\footnote
{Washington Double Star Catalogue (WDS), \citet{WDS},
http://ad.usno.navy.mil/wds/orb6.html}).

The close, eclipsing binary producing the
light variations (from now on: stars 1 and 2) is
the brighter component of the visual pair. The observed separation of
the visual components was $0.1 - 0.2$
arcsec during the available astrometric observations \citep{WDS}
which is much less than the typical seeing at the DDO of 1 -- 3 arcsec;
therefore, the spectra of both components were observed simultaneously.
We discovered that the broadening functions
(Figures~\ref{fig4} and \ref{fig5})
show an occasional splitting of the third-component peak, indicating
that it is in fact a close, non-eclipsing binary (from now on: stars 3
and 4) with a very strongly elliptical orbit. Thus, the system is a
hierarchical quadruple with both components of the visual 113 year period
system being double-line (SB2) close binaries.
Our estimates of the combined brightness of the third and fourth
components are $L_{34}/(L_1 + L_2) = 0.35 \pm 0.02$ for the
spectral region around 5184~\AA\ and $L_{34}/(L_1+L_2) = 0.33 \pm 0.02$
at 6400~\AA\ during the maximum light of the closer pair.

The RV data for the close binary
were handled in the standard way, by first removing the peaks
(by preliminary fitting of Gaussian profiles) of the second binary
and then measuring the positions of the RV peaks for the close pair.
The novelty here is that instead of Gaussians, as in previous papers,
we used the double rotational profiles for the close pair (Section~\ref{rot}).
The results indicate that the brighter component of ET~Boo is a semi-detached
or more likely a detached binary with a relatively large mass ratio, $q = 0.884$.

Only a small fraction of the available BF's show splitting of
the visual companion peaks (components 3 and 4); this property
gave us a hint of a highly eccentric orbit, but also crucially
helped in finding the orbital period.
Because these stars have very similar brightness, the period
ambiguity was resolved by consideration of the RV
differences between the components. A preliminary orbital period found
by trigonometric polynomial fitting to the data was
later refined by the spectroscopic orbit solution
(Table~\ref{tab4}, Fig.~\ref{fig6}) to give 31.521 days. The systemic
velocity of the second binary $V_0 = -24.15 \pm 0.44$ km~s$^{-1}$ is
close to the systemic velocity of the close pair $V_0 = -23.52 \pm 0.52$
km~s$^{-1}$ confirming the physical association of the two
binaries. Since the
orbital period of the wide pair is 113 years, no orbital motion can
be expected to be detected in the 5 years of our observations.

\subsection{TX~Cnc}
\label{txcnc}

The photometric variability of TX~Cnc, an apparent member of
the Praesepe open cluster, was first announced by \citet{haff1937}.
\citet{whel1973} were the first to obtain good simultaneous fit
to the photometric and spectroscopic data of the system on the assumption
of the Roche model. A preliminary analysis of
the DDO observations was published in a PhD thesis \citep{blake2002}.
The RVs used there have been later re-determined
using the rotational profile fitting.
Surprisingly, the broadening functions
show a shape of a practically detached binary (see Fig~\ref{fig4}),
although the orbital velocities are quite typical for a contact,
W-type system.

The spectroscopic elements of TX~Cnc determined
by \citet{whel1973} were
$V_0$ = 26.6 $\pm$ 3 km~s$^{-1}$,
$K_1$ = 117.3$\pm$3 km~s$^{-1}$, $K_2$ = 189.8 $\pm$ 4 km~s$^{-1}$,
giving $q$ = 0.62. A later RV solution published
by \citet{lean1983},
$V_0  = 29 \pm 6$ km~s$^{-1}$, $K_1 = 96 \pm 8$ km~s$^{-1}$
and $K_2 = 181 \pm 11$ km~s$^{-1}$,
was based on only 8 photographic spectra.
Our spectroscopic elements suggest a smaller
mass ratio $q = 0.455 \pm 0.011$ than derived before and a larger
total mass, $(M_1 + M_2) \sin^3 i = 1.330 \pm 0.012 M_\odot$; both
changes are rather characteristic for an improved quality of the
spectroscopic observations.

TX~Cnc is of particular interest for our understanding of the
evolution of contact binaries because it appears to belong to
Praesepe, which is one of the youngest (900 Myr) open clusters
containing such binaries \citep{ruc98}. All indications point at
an advanced age of contact binaries
so that confirmation of the membership of TX~Cnc to Praesepe may
provide a much needed lower limit on the time needed to form such
binaries. Unfortunately, the system's parallax was not
measured during the Hipparcos mission so that the membership
must be judged by less direct means. Our radial velocity data giving
$V_0 = 34.0 \pm 0.5$ km~s$^{-1}$, are fully consistent with the
mean velocity of the Praesepe cluster,
$V = 34.53 \pm 0.12$, and the velocity dispersion of its spectroscopic
binaries, $\sigma = 0.40$ km~s$^{-1}$ \citep{merm1999}.

In the PPM Catalogue, \citet{roba1988} assign TX Cnc a parallax of $5.21
\pm 0.79$ mas.
Hipparcos astrometry of Praesepe analyzed by \citet{lee1999} gives
a cluster parallax of $5.32 \pm 0.37$ mas, i.e. a distance modulus
of $m-M = 6.37 \pm 0.15$. With the maximum brightness of
TX~Cnc of $V_{max} = 10.0$, we get $M_V = 3.63 \pm 0.15$
which is in perfect agreement with the absolute magnitude estimated
from the \citet{rd1997} calibration giving $M_V = 3.60$, for  assumed
$(B-V)_0$ = 0.54  corresponding to  spectral type F8V.

This excellent agreement in radial velocity, parallax and luminosity
distance is supported by proper motion data. The careful photographic study by
\citet{js91} showed that TX Cnc has a proper motion of
$(\mu_{x}=-4.4 \pm 3.2, \mu_{y}= 0.5
\pm 1.4)$ mas yr$^{-1}$ relative to the centre of motion of Praesepe,
leading \citet{js91} to assign a 99\% probability that TX Cnc belongs to
Praesepe. The velocity dispersion contributes less than 1 mas yr$^{-1}$. The
Tycho-2 work of \citet{Tycho2} yields $(\mu_{\alpha} \cos \delta = -36.2
\pm 1.2, \mu_{\delta}= -11.2 \pm 1.3)$
mas yr$^{-1}$ for the absolute proper motion of TX Cnc, compared with the
mean centre of mass motion of $(-35.7 \pm 0.4, -12.7 \pm 0.3)$ mas yr$^{-1}$
found by \citet{lee1999}. We can therefore be very confident that TX Cnc is
a member of Praesepe, and hence it is an important system for testing
theories of contact binary formation and evolution.

\subsection{V1073~Cyg}
\label{v1073cyg}

The bright A-type contact binary V1073~Cyg was a subject of several
photometric studies \citep{sezer1996,morris2000}. There exist also
two spectroscopic investigations: \citet{fitz1964} published a
spectroscopic orbit for the primary component with a marginal detection of
the secondary, with $V_0 = -8\pm3$ km~s$^{-1}$,
$K_1 = 66 \pm 4$ km~s$^{-1}$ and the mass ratio of $q$ = 0.34. The author found
a small eccentricity of the orbit, $e = 0.115 \pm 0.053$,
which cannot be significant when one applies arguments of
\citet{Lucy71,Lucy73} on statistics of eccentricity determinations.

\citet{ahn1992} obtained an apparently more reliable, circular orbit
solution for both components, with the resulting spectroscopic elements,
$V_0 = -0.8 \pm 1.1$ km~s$^{-1}$, $K_1 = 66.7 \pm 1.3$ km~s$^{-1}$
and $K_2 = 210.2 \pm 1.2$ km~s$^{-1}$.
Our results, $V_0 = -6.85 \pm 0.50$ km~s$^{-1}$, $K_1 = 65.53 \pm 0.64$
km~s$^{-1}$ and $K_2 = 218.9 \pm 1.5$ km~s$^{-1}$, are superior to the previous
ones thanks to the BF extraction technique and the rotational profile fitting.
The relatively large formal $rms$ errors are mainly caused by the
simple sine curve solution used by us. When combined with a high-precision
light curve, the BF modeling can provide high-quality absolute parameters of the
system.

Our spectral type estimate, F0V,  is much later than original
classification of \citet{fitz1964} who estimated the
spectral type as A3Vm; it
confirms the classification of \citet{hill1975}.
The Tycho-2 color $B-V$ = 0.375 and our spectral
type indicate a non-negligible amount of reddening. The Hipparcos parallax,
$\pi = 5.44 \pm 0.95$ mas is not sufficiently precise
to draw conclusions on the physical properties of the system.

\subsection{HL~Dra}
\label{hldra}

Variability of HL~Dra was detected during
the Hipparcos mission. The system was
classified as a $\beta$~Lyrae eclipsing binary with the orbital period of
0.944276 days. The primary component is of the A5
spectral type. No ground-based photometric study of the system has been
published yet. Also no recent minima after the Hipparcos mission are
available. Our time of the spectroscopic conjunction shows only a small
shift (+0.0047 days) with respect to the time predicted by the original
Hipparcos ephemeris so that the orbital period of the system appears to
be very stable.

We have not been able to detect spectral signatures of the secondary component
in our data. The broadening functions close to quadratures
show only small humps on both sides of the primary peak which
cannot be identified with the secondary component because
they do not show any orbital motion.
The system is clearly a detached or semi-detached pair with a low
luminosity secondary component.

HL~Dra was observed during two seasons in two different wavelength
regions, in 2004 at 5184~\AA~ and in 2005 at
6290~\AA. The latter dataset is of relatively poor
quality due to the low number and weakness of spectral lines in the
red spectral region of an early type star. The orbital single-line solutions
resulting from the two datasets are in a good accord except
for the center of mass velocity of $V_0=-29.3 \pm 0.4$ km~s$^{-1}$
for the 2004 data and $V_0=-36.5 \pm 0.6$ km~s$^{-1}$
for the 2005 data. The shift is well outside the formal errors and may
be caused by a motion of the eclipsing pair around the barycenter with
a third body. The 2004 data are of much better quality so the orbital
parameters listed in Table~\ref{tab1} correspond to this dataset.

Our new spectral type determination is slightly later, A6V,
than previously published. The Tycho-2 color
index $(B-V) = 0.222$ corresponds to the A8V spectral type so that the
reddening appears to be small.

\subsection{AK~Her}
\label{AK Her}

AK~Her is W UMa-type contact binary discovered by Metcalf (see
\citet{pick1917}). It is the brighter component in the visual binary
ADS~10408. The companion, located at a separation of
4.2 arcsec, is 3.5 mag fainter than AK~Her at its maximum light. The
position angle 323\degr~ is almost perpendicular to our fixed, E--W
spectrograph slit so that this component
was not detectable in the broadening functions.

The system is known to show
a cyclic variation in the moments of eclipses which is
probably caused by the light-time effect induced by an
undetected companion on an orbit of about 57 years
\citep{awad2004}. The perturbing star cannot be identified with the known
visual companion and must be much closer to the
binary. The complex multiplicity of the system is supported
by the Hipparcos astrometry by the following: (i)~The
system shows a stochastic
astrometric solution (the X flag in the catalog field H59);
(ii)~It is suspected not to be single (the S flag in H61),
(iii)~The trigonometric parallax of $10.47 \pm 2.77$ mas
has a much too large error for the brightness of the system.
Our individual spectroscopic data do not show any contribution
from this putative third (or rather fourth)
component. It is possible that such a companion
will be seen in a detailed analysis of
averaged spectra \citep{dangelo2006}, but this approach
is outside the scope of the present paper.

Our RV solution is the first to treat the star as a double-lined
binary system (SB2).
\citet{sanf1934} observed the RV curve of the primary
component and determined $f(m)$ = 0.0208 M$_\odot$. His RV
solution ($V_0 = -13$ km~s$^{-1}$, $K_1 = 79$
km~s$^{-1}$) is fairly consistent with our solution.

Our spectral classification gives an earlier spectral
type for the system, F4V, than previously discussed, F8V. It is not
fully consistent with the Tycho-2 color index $(B-V) = 0.490$ and
implies some interstellar reddening.

\subsection{VW~LMi}
\label{vwlmi}

The photometric variability of VW~LMi was found by the Hipparcos
mission. It was classified in the Hipparcos catalogue \citep{hip}
as a W~UMa type eclipsing binary with an orbital period 0.477547
days. The first photometric observations of the system were published
by \citet{dumi2000}. Later the light curve of the system was
analyzed by \citet{dumi2003}, who found the mass ratio
$q_{ph}$ = 0.395 and
inclination $i$ = 72.4\degr; we show later that these values
are incorrect as they do not take into account the presence of
the relatively bright binary companion.

VW~LMi has been observed spectroscopically at DDO since 1998.
It was realized from the beginning that the system is a quadruple one,
consisting of two spectroscopic binaries. While the eclipsing pair
can be identified with the short-period contact binary,
the second spectroscopic binary is a detached one with the period
of about 7.9 days (see below how we arrived at the
more exact value). The light contribution of the second spectroscopic
binary at the maximum brightness of the contact pair is
$L_{34}/(L_1 + L_2) = 0.42$.

The study of quadruple system VW~LMi is complicated by the mutual
orbital motion of both binaries so that changes in the respective
$V_0$ values cannot be neglected. Another complication is the
similar brightness of components of the second, non-eclipsing
binary making derivation of its orbital period very difficult.
We worked first with RV differences of the components
of the second pair to find its orbital period. Trigonometric polynomial fits
to the data led to the orbital period $P_{34}$ = 7.9305 days, which explained
all data very well. An attempt to find the orbit of the contact pair
(after removing the contribution of the second binary)
resulted in poor quality of the
spectroscopic orbit. In fact, the residuals from
both preliminary orbits showed a clear anti-correlation
between the velocities of the contact pair (components 1 and 2)
and of the non-eclipsing binary (components 3 and 4) indicating the mutual
orbital motion of the two binaries. The period analysis
revealed only one feasible period of about 355 days, hence close to one
year. Since the data span 7 years and the observing season
for VW~LMi is from late November to mid May, the phase coverage of the
mutual orbit is partial and has gaps.

A further improvement of all three orbits was achieved by simultaneous fits to
all four datasets of the RVs of the form:
\[V_{i} = V_0 + (-1)^i K_i [e_{j} \cos \omega_{j} + \cos (\omega_{j}+\nu_{j})]
             + (-1)^j K_{2j-1,2j} [e_3 \cos \omega_3 + \cos (\omega_3 +\nu_3)]\]
where $V_0$ is the mass-center velocity of whole quadruple system, $K_i$ are the
respective velocity semi-amplitudes of individual components, $e_j$ is orbital
eccentricity, $\omega_j$ is longitude of the periastron and $\nu_j$
is the true anomaly. The index $i$ corresponds to the component number
($i = 1 - 4$), while the index $j$ takes the value of
1 for the contact binary, 2 for the detached
binary and 3 for the mutual orbit of the two systems.
Thus, for the components of the contact binary, $j=1$ and $i=1,2$ while
for the components of the detached binary, $j=2$ and $i=3,4$.
$K_{2j-1,2j}$ for $j$=1 should be read as $K_{12}$ while for $j=2$ as $K_{34}$
where
$K_{12}$ and $K_{34}$ are the semi-amplitudes of the RV changes
of the mass centers of the contact and the detached binary, respectively.

All results of the simultaneous fits are presented in Table~\ref{tab5}
while the orbital elements of the contact pair are also given with the remaining
binaries of this study in Table~\ref{tab2}.
For simplicity, all measurements were assigned the same weight
although the velocities for the detached pair were determined by
the Gaussian fits while those of contact pair by the rotational
profile fits. The sine curve fits to the data for the contact
binary, corrected for the motion on the outer orbit, are shown
in Fig.~\ref{fig2}.  While the secondary component is
usually not blended with the peaks of the second spectroscopic binary,
the primary of the contact pair is always visible projected against
the ``background'' of the profiles of the 3rd and 4th component.
This circumstance caused an enhanced
scatter of the velocities of the primary component.

The corrected RVs of the second binary with the corresponding fits
are plotted in Fig.~\ref{fig7}. The final orbital period for this binary is
7.93044 days and the orbit is nearly circular. The velocities of all four
components corrected for the corresponding orbital motions in the inner orbits
and their best fits are plotted in Fig.~\ref{fig8}. These residual RVs
represent the orbital motion of the mass centers of both binaries.

Because the outer orbit has a relatively short period of 355 days,
it is of interest to inquire into the mutual orientation of the
three orbits. This can be estimated from projected masses of the
components in a sort of a ``bootstrap'' process started with the
derived inclination of the contact, eclipsing system. A preliminary
solution of unpublished photoelectric data obtained at the Star\'a
Lesn\'a Observatory of the Astronomical Institute of the Slovak
Academy of Sciences were used to estimate inclination angle of the
eclipsing pair. Fixing the third light at
$L_{34}/(L_1 + L_2) = 0.42$ (see above) and
using the spectroscopic mass ratio of $q=0.42$ led to the
inclination angle $i_{12} = 80.1^\circ \pm 0.2^\circ$. This is, as
expected, a much larger inclination than the one obtained by
\citet{dumi2003} ($i = 72.4^\circ$) without an assumption of the
third light. Using our estimate of the inclination angle and the
projected total mass of the contact pair $(M_1+M_2) \sin i_{12} =
2.28\, M_\odot$, we obtain $(M_1 + M_2) = 2.39\, M_\odot$. The
outer, 355 day orbit defines the mass ratio for the two pairs,
$(M_1+M_2)/(M_3+M_4) = 1.09$. Therefore, the true (not the
projected) mass of the second spectroscopic binary is $(M_3+M_4) =
2.19\, M_\odot$. Using the projected mass $(M_3+M_4) \sin i_{34} =
1.79 \pm 0.03\, M_\odot$, we estimate the inclination of the orbit
of the second pair to be about $69^\circ$. The outer, 355 day
orbit is even less inclined to the sky since with
$(M_1+M_2+M_3+M_4) = 4.58\, M_\odot$ and the projected total mass
of only $2.67\, M_\odot$, so that we obtain $i_{12-34} = 57 \degr$.
Obviously, we could not determine if these values are all in the
same sense or are complements to $180^\circ$; a determination
of the sense of the revolution could only come from interfermetric
observations.

The non-eclipsing, detached binary,
with $(M_3+M_4) = 2.19\, M_\odot$, is
composed of two almost identical ($q_{34} = M_3/M_4 = K_4/K_3 =
0.980 \pm 0.017$), most probably main sequence stars. Their masses
correspond to about F9V -- G0V and their similarity is also
reflected in the luminosity ratio $l_3/l_4 \approx$ 1.04. The evolutionary
status of components can be guessed from a comparison of
their rotational and orbital velocities. If we assume a
synchronous rotation and take rotational velocities of the
components estimated from Gaussian profile fits
to be about 12 km~s$^{-1}$ and semi-amplitudes of
RVs about 60 km~s$^{-1}$, we see that fractional radii of
components are $r/a<0.2$. With the semi-major axis of the absolute orbit of
about 10 R$_\odot$, their radii are $<2 R_\odot$.
The similar spectral type of all components of the
quadruple system of  VW~LMi has resulted in a very good
quality of extracted BF's, as can be judged in Figure~\ref{fig4}.

It is interesting to note that multiplicity of VW~LMi was not
identified astrometrically during the Hipparcos mission in spite
of the relative proximity of the system at $\pi = 8.04 \pm 0.90$ mas.
This is probably caused by the orbital period of the
two pairs around each other being close
to one year thus mimicking the parallactic motion.

Because of the small size of the mutual (355 day)
orbit of only 0.62 AU, chances of resolving the two astrometric components
are rather small even with advanced techniques because the
expected maximum angular separation will be only 10 mas.
The situation is little bit more optimistic with the expected
light-time effect of the eclipse timing of the contact binary
as the expected full amplitude is about 0.0074 days
which should be relatively easy to detect
with the observed photometric amplitude of 0.42 mag.

\subsection{V566~Oph}
\label{v566oph}

The W~UMa-type binary V566~Oph was discovered by \citet{hoff1935}.
The system is bright ($V_{max}$ = 7.46) and therefore it was a
subject of numreous previous photometric
(for references, see \citet{twig1979}) and spectroscopic observations.
An interval of constant light observed during the
secondary eclipse indicates the A-type.
\citet{MD72b} published the first light curve analysis of V566~Oph based on the
Roche model. The total eclipses permitted to determine
reliable geometric elements: the fill-out $F=1.25 \pm 0.05$ ($f$ = 0.25),
mass ratio $q = 0.238 \pm 0.005$ and inclination of $i = 80\degr \pm 2\degr$.

There exist three previous spectroscopic studies of the system. Two of these are
based on photographic observations \citep{hear1965,mclean83}
and a more recent one is based on the Reticon data
\citep{hill1989}. The latter study used direct fitting of the synthetic profiles
to the CCF functions. The spectroscopic elements obtained in his study,
$V_0$ = 38.5$\pm$1.1 km~s$^{-1}$, $K_1$ = 72.6$\pm$1.5 km~s$^{-1}$ and
$K_2$ = 272.9$\pm$1.3 km~s$^{-1}$,
are practically within the errors of our results,
$V_0$ = 37.3$\pm$0.5 km~s$^{-1}$, $K_1$ = 71.1$\pm$0.7 km~s$^{-1}$ and
$K_2$ = 270.1$\pm$1.1 km~s$^{-1}$. The current improvement of the orbit
is mainly due to the use of the BF extraction technique and of
the rotational-profile fitting.
The orbit can be still improved by taking the proximity effects into account.
Our new determination of the mass ratio, $q  = 0.263 \pm 0.012$ is
in a moderately good agreement with photometric mass ratio of \citet{MD72b},
confirming the utility of the photometric approach to systems with total eclipses.

The orbital period of the system is rather unstable. In spite of the possible
light-time effect orbit found by \citet{priruc2006}, we did not find any traces
of the third component in the extracted BFs.

V566~Oph is a relatively nearby system with a good Hipparcos parallax,
$\pi  = 13.98 \pm 1.11$ mas. The absolute magnitude determined using
the calibration of \citet{rd1997}, $M_V = 3.07$, using $(B-V)=0.406$
from the Tycho-2 catalogue \citep{Tycho2} is in good agreement with
the visual absolute magnitude determined from the Hipparcos
parallax, $M_V = 3.19 \pm 0.17$.
Our new spectral classification, F4V, indicates a slightly
later spectral type than that F2V found by \citet{hill1989}.

\subsection{TV~UMi}
\label{tvumi}

TV~UMi is another discovery of the Hipparcos mission. The system was
classified as a $\beta$~Lyrae eclipsing binary with orbital period of
0.415546 days, although the classification was obviously
complicated by the low amplitude of the light
variations of only about 0.08 mag. The eclipses are very wide and of
almost the same depth. For that reason we suspected that the variability
is caused by a contact binary of the W~UMa-type.

Prolonged spectroscopic observations of TV~UMi at the DDO showed
that the system is a quadruple one and consists of two
spectroscopic binaries. The second close binary in TV~UMi is
almost as bright as the contact pair, but its components are
difficult to analyze because of the strong eccentricity of the
orbit and the very short duration of periastron passages when the
spectral signatures could be potentially resolved. In fact, the
components of the second pair could not be separated in most of
our spectra; such a separation took place on only
three occasions.
The largest observed separation of the components of the
second binary on June 8, 2001 of more than 100 km~s$^{-1}$
had to be close to periastron passage.
One of the BF's from that night is included in Fig.~\ref{fig3}. During
this event, the stronger component had a more negative RV. A period
analysis of line separations indicates a 31.2-days orbital
periodicity of the second pair. The reliable determination of the
orbital parameters and confirmation of the preliminary orbital
period would require more observations, preferably with a
larger spectroscopic resolution to
separate the components even outside the periastron passages. The TV~UMi
system resembles ET~Boo in that its companion binary also has an
eccentric orbit with a similar orbital period. Our RV observations
cover a shorter time base and are less numerous than in the case
of VW~LMi so that we have not been able
to find the mutual orbital motion of the binaries. During intervals
when we observed the narrow blend of the peaks of the third and fourth
components, the combined RV was about $-15$ km~s$^{-1}$, slightly less than
systemic velocity of the contact binary ($-9.7$ km~s$^{-1}$). This
indicates a possible slow orbital motion of both pairs.

The light contribution of the second binary is large,
$L_{34}/(L_1+L_2) = 0.90$. As observed during the periastron passage
on June 8, 2001, its components have slightly unequal brightness:
$L_3/(L_1+L_2) = 0.58$ and $L_4/(L_1+L_2) = 0.32$.
Using the measured RVs of the third
and fourth components during the periastron passage,
$RV_3 = -67.94 \pm 1.13$ km~s$^{-1}$ and $RV_4 =
37.21 \pm 1.03$ km~s$^{-1}$,  and the mean RV of the
blend of the two components giving the approximate systemic velocity of
the second pair, $-15$ km~s$^{-1}$, we see that the
mass ratio of the second binary is close to unity.
The observed photometric amplitude of the contact system,
$\Delta m = 0.083$, when corrected for light contribution of
the companion binary, remains small at $\Delta m$ = 0.163.

It is interesting to note that the system was not detected nor
even suspected as a multiple one from the Hipparcos astrometric
data. Chances to resolve both binaries by direct imaging
are higher than in
VW~LMi because no rapid orbital motion of the contact binary was
observed indicating a longer orbital period and thus a larger separation
of both binaries. According to the Hipparcos astrometry TV~UMi is
relatively nearby, with $\pi = 7.64 \pm 0.78$ mas.

Clearly, the current data for the second close binary system
in TV~UMi are inadequate for determination of full
orbital parameters for the whole system. Such a determination
would require a long-term monitoring program with one or two spectra
obtained per night over a period of few months.

\subsection{AG~Vir}
\label{agvir}

Variability of AG~Vir was discovered by \citet{guthprag1929}.
Since then it was a subject to several photometric investigations
(for references, see \citet{bell1990}). The system is very similar to
DU~Boo (Section~\ref{duboo}) in that
the first light maximum is always brighter of the two by
about 0.08 mag. A simultaneous photometric and spectroscopic
analysis of \citet{bell1990} showed that a reliable
determination of the geometrical elements was complicated by the strong
asymmetry of the light curve. The system is of the A-type
with an undeterminable degree of contact (could be marginal or
deep). The spectral type is A7V (our new
determination suggests A5V), so that the observed photospheric brightness
inhomogeneities may be quite different from the solar-type dark spots.

The period analysis of \citet{blancat1970} indicated the
presence of a light-time effect in the eclipse timing
caused by a third component on a 40 years orbit. However, later
observations did not confirm any cyclic behavior, although the observed
times of minima do show a large scatter. This can be interpreted
either by the light-curve variations caused by the presence of
surface spots or by a light-time effect caused by a body on
a short-period orbit. In fact, our broadening functions
do show a well defined signature of the
third component with $L_3/(L_1 + L_2) \approx 0.05$
(see Fig.~\ref{fig9}). However, this star has a rather different
RV ($+2.9 \pm 2.5$ km~s$^{-1}$) from the systemic velocity
of the contact binary ($-10.99 \pm 0.82$ km~s$^{-1}$).
Unfortunately, the uncertainties of the RVs are quite large and only our
better-quality 2005 spectra (as used in this paper)
show the presence of this component. A close inspection of the published
cross-correlation functions of \citet{bell1990} reveals a
marginally defined presence of a third component close to
the systemic velocity of the binary, but -- as expected --
the definition of the CCF's was inferior to that of the BF's.

The Hipparcos parallax ($1.33 \pm 1.18$ mas) is too small and
inconsistent with the estimated absolute magnitude for a A5V star,
$M_V = +2.1$, and the observed maximum visual magnitude,
$V_{max} = 8.50$. It is possible that this inconsistency
is caused by a transverse motion of the eclipsing pair
around the third component.

\section{Measurements of radial velocities using rotational profiles}
\label{rot}

While the previous papers of this series (Papers
I--VI and VIII--X) reported
RV measurements obtained by Gaussian fits to the
extracted broadening functions, in this paper we use a novel
technique of extracting RVs by rotational profile fits.
While neither Gaussians nor rotational profiles can replace the
full modeling of broadening functions which will
hopefully take place one day, the rotational profiles are
equally simple to implement as the Gaussians, but offer
an improvement in the quality of the RV
measurements with a much better convergence to the final
result and noticeably smaller random errors.

The broadening function of a rigidly rotating star \citep{gray1976}
is described by four parameters:
the overall strength or the amplitude of the BF, $a_0$,
the central velocity, $a_1$ (identified with the light
centroid velocity of the star), and the half-width
$a_2 = V \sin i$; an additional parameter is a vertical
background displacement, $a_3$, which can be usually traced to a different
continuum and pseudo-continuum levels
for the standard and program stars during the spectral
normalization step. For a double-peaked profile, the three
first parameters ($a_0$ to $a_2$)
appear twice so that the fit involves 7 unknowns.
By using auxiliary quantities, $c_1 = 1-u$ and
$c_2=\pi u/4$, the profile can be written as:

\begin{equation}
I=a_0+a_1 \left[ c_1 \sqrt{1-(x-a_1)^2/a_2^2}
+ c_2 \left(1-(x-a_1)^2/a_2^2\right)\right] /(c_1+c_2).
\end{equation}

The profile depends only slightly on the limb darkening coefficient, $u$.
We assumed $u=0.7$ in our measurements.
As opposed to the Gaussian profile, the rotational
profile (see an example in Fig.~\ref{fig10}) has rather steep sides
and -- by definition -- is exactly zero for velocities larger than
projected equatorial velocity $V_{rot} \sin i$. These properties
are crucial to the improvement in the determination of the
parameter $a_1$, the light centroid for each component.

Although the rotational profiles represent the BF profiles of
double-line binaries much better than Gaussians,
a practical application may encounter the same complications:
\begin{itemize}
\item If the secondary shows a very faint peak in the BF, its
width must be usually fixed at a reasonable value to improve
the stability of the solution and to give consistent results;
\item The BF profiles for contact binaries are rather different
from those of single, rigidly rotating stars, particularly in the ``neck''
region between the stars where an additional light is present.
Obviously, these asymmetric deviations cannot be fitted by
rotational profiles nor by Gaussians. Normally, this
leads to under-estimates of the velocity semi-amplitudes.
\end{itemize}
In the case of close binary stars, the combined rotational profiles
can be applied to broadening functions strictly only outside the
eclipses as a proper representation of the data should involve inclusion of the
eclipse and proximity effects \citep{ruc92}. However, a simple upper
envelope of the two individual profiles works well even
during partial eclipse phases. We found that
the double rotational profiles converge faster to the final results and
describe the data much better than the Gaussians. This is well
illustrated in the case of TX~Cnc, where a preliminary orbit defined by
RVs obtained through a Gaussian fits
had almost twice as large standard errors of the spectroscopic elements.
While $rms$ errors for the velocities
derived from the Gaussian fits were $\sigma_1$ = 8.4
km~s$^{-1}$ and $\sigma_2$ = 8.3 km~s$^{-1}$, the errors for the
RV's derived by rotational profile fitting are
$\sigma_1$ = 4.1 km~s$^{-1}$ and $\sigma_2$ = 5.7 km~s$^{-1}$.

\section{SUMMARY}

With the new ten short-period binaries, this paper brings the
number of the systems studied at the David Dunlap Observatory to
a round number of one hundred. The systems presented in this paper
include three quadruple systems for which we have been
collecting data for several years in a hope of being able to study
variability of RVs on times scales ranging from
a fraction of the day to several years. This has been achieved for
ET~Boo and VW~LMi where we can say much about all components of
these hierarchical binaries. ET~Boo is a known visual binary
with the period of 113 years with each component being a close
binary. VW~LMi is a particularly
interesting system with the period of mutual revolution of both
binaries of only 355 days. Starting with a preliminary
photometric solution of the light curve of VW~LMi which gave the
orbital inclination of the close binary, we were able to determine
the orbital inclinations of all involved binaries in this
system and thus to derive masses of all components. We have been
less successful for the quadruple system TV~UMi where the
second pair requires a very prolonged monitoring for analysis
of the 31.2 day orbit of the second pair.

We have found that AG~Vir appears to be a triple system,
although there is inconsistency in the velocity of
its companion. For AK~Her, we were able to obtain data free of
contamination from
the known third component; there are indications that this binary
has another faint companion causing the light-time effect in
eclipse timing. The systems DU~Boo, TX~Cnc, V1073~Cyg, V566~Oph
are relatively mundane double-lined contact binaries
while HL~Dra is a single-lined binary of an unknown variability type.

All RVs for close binaries analyzed in this paper have been
determined by using a novel technique of rotational profile
fitting to the broadening functions. This technique, while
still not perfect in reproducing asymmetries and
intricacies of the real BF's, is much advantageous and
accurate than the Gaussian fitting previously used in our
studies.

\acknowledgements

We express our thanks to Christopher Capobianco, Kosmas Gazeas,
Panos Niarchos, Matt Rock, Piotr Rogoziecki, and Greg Stachowski,
for their contribution in collecting the observations.  

Support from the Natural Sciences and Engineering Council of Canada
to SMR and SWM and from the Polish Science Committee (KBN
grants PO3D~006~22 and P03D~003~24) to WO and RP
is acknowledged with gratitude. The travel of TP to
Canada has been supported by a IAU Commission 46 travel grant and a
Slovak Academy of Sciences VEGA grant 4014.
TP appreciates the hospitality and support of the local staff
during his stay at DDO. M. Blake acknowledges support through an NSERC grant
to C. T. Bolton.

The research made use of the SIMBAD database, operated at the CDS,
Strasbourg, France and accessible through the Canadian
Astronomy Data Centre, which is operated by the Herzberg Institute of
Astrophysics, National Research Council of Canada.
This research made also use of the Washington Double Star (WDS)
Catalog maintained at the U.S. Naval Observatory.

\clearpage

\noindent
Captions to figures:

\bigskip

\figcaption[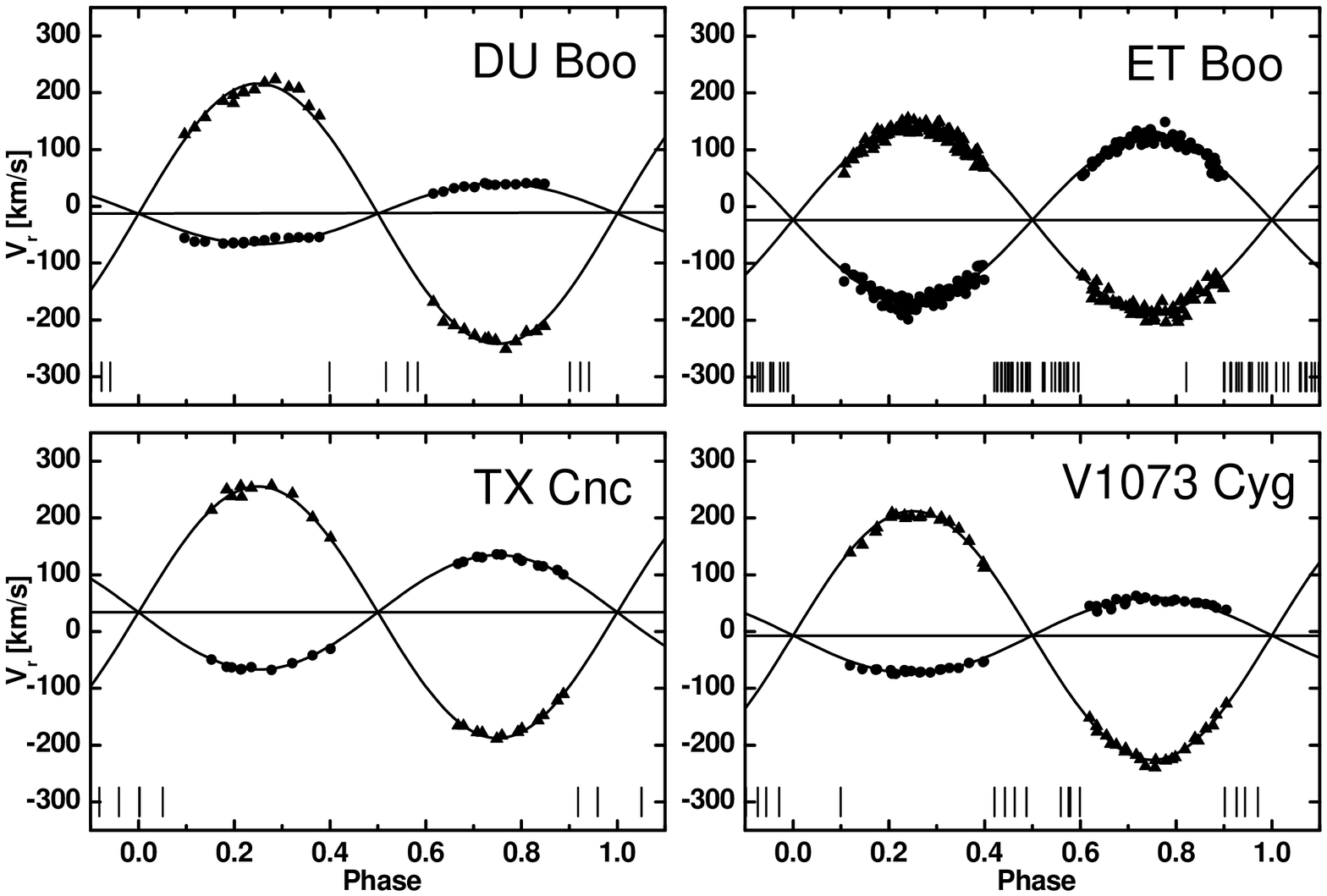] {\label{fig1} Radial velocities of the
systems DU~Boo, ET~Boo, TX~Cnc and V1073~Cyg are plotted in
individual panels versus the orbital phases. The lines give the
respective circular-orbit (sine-curve) fits to the RVs.
ET~Boo is quadruple system composed of detached
eclipsing pair (wide profiles) and detached non-eclipsing pair with a
much longer orbital period $P \approx$ 31.5 days (narrow peaks).
DU~Boo, TX~Cnc and V1073~Cyg are contact binaries. The circles and
triangles in this and the next two figures correspond to
components with velocities $V_1$ and $V_2$, as listed in
Table~\ref{tab1}, respectively. The component eclipsed at the
minimum corresponding to $T_0$ (as given in Table~\ref{tab2}) is
the one which shows negative velocities for the phase interval
$0.0 - 0.5$. Short marks in the lower parts of the panels show
phases of available observations which were not used in the
solutions because of the blending of lines. All panels have the
same horizontal range, $-500$ to $+500$ km~s$^{-1}$. }

\figcaption[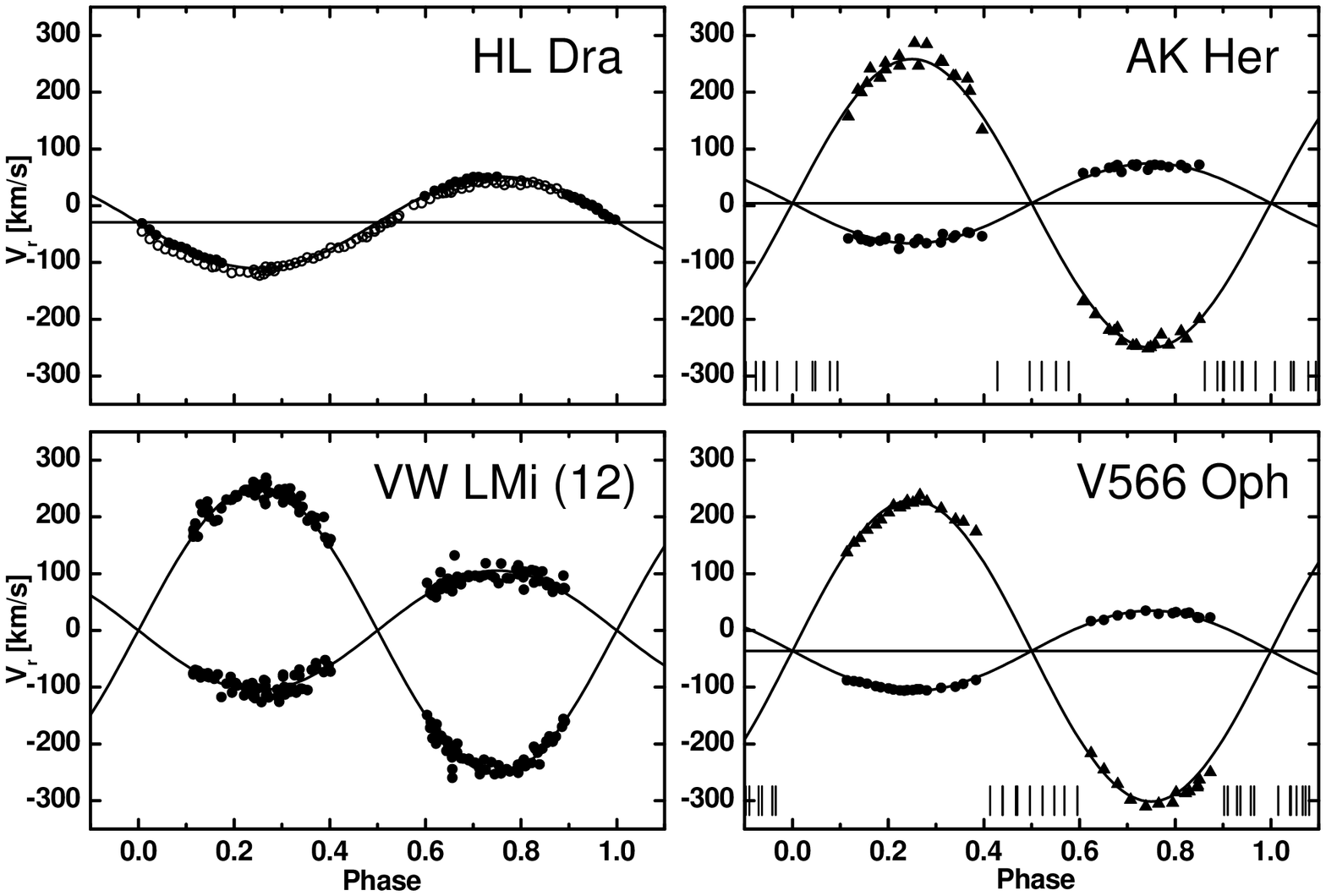] {\label{fig2} The same as for
Figure~\ref{fig1}, but for HL~Dra, AK~Her, VW~LMi, and V566~Oph.
HL~Dra is a detached single-lined binary while AK~Her, and
V566~Oph are contact binaries. VW~LMi is a quadruple systems
containing the eclipsing contact binary and a detached
non-eclipsing binary with a 7.93 days orbit. The study and
interpretation of broadening functions is complicated by the fact
that both binaries revolve in a relatively tight orbit with $P
\approx 355$ days. We show additional data for VW~LMi in
Figures~\ref{fig7} -- \ref{fig8}. }

\figcaption[f3.ps] {\label{fig3} The same as for
Figures~\ref{fig1} and \ref{fig2}, but for the two remaining systems
TV~UMi and AG~Vir. While AG~Vir is contact binary containing a third
component, TV~UMi is quadruple system containing a contact binary
and a detached non-eclipsing binary with a period of about 31.2 days.}

\figcaption[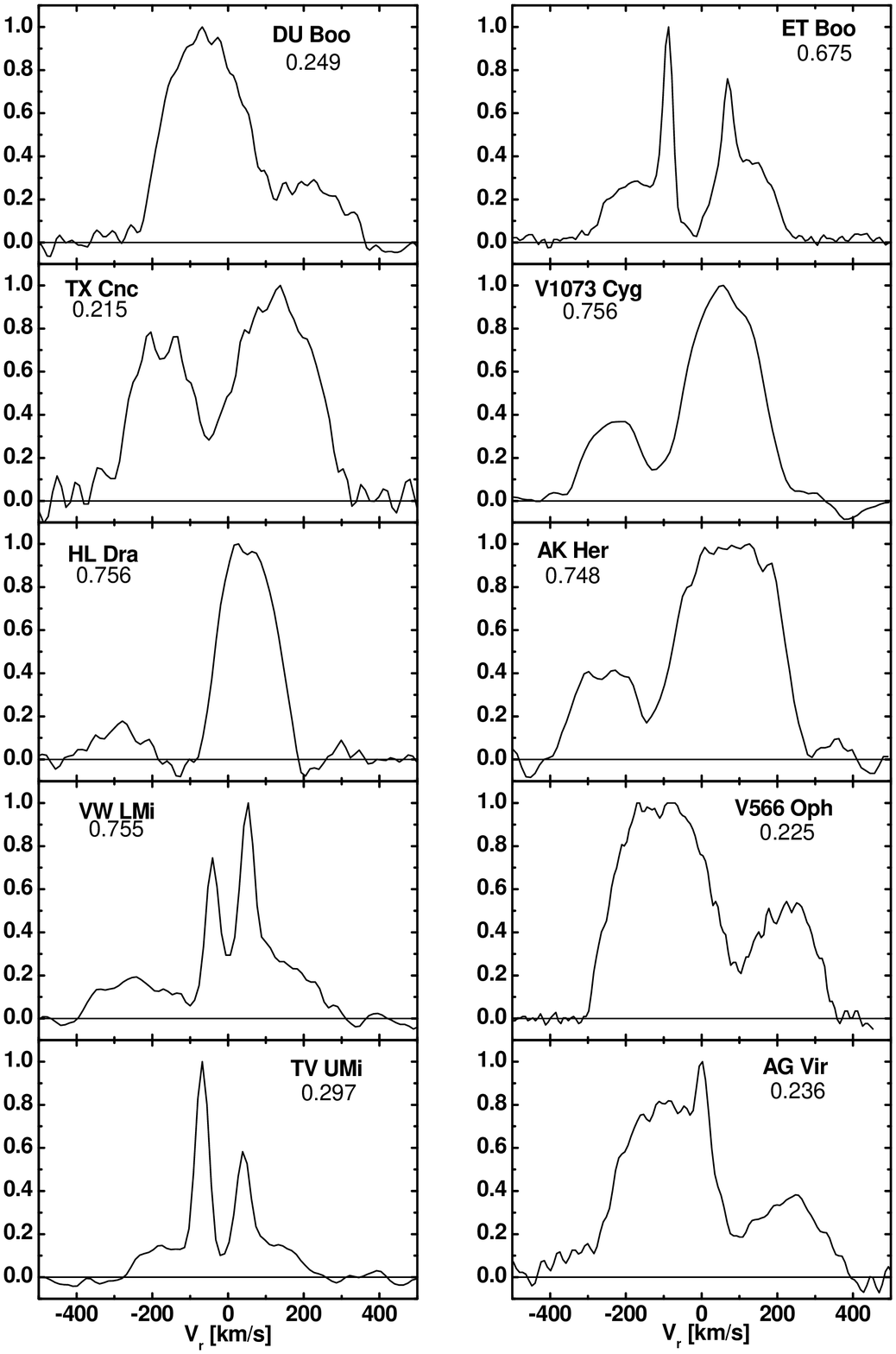] {\label{fig4}
The broadening functions (BF's) for all ten systems of this
group, selected for phases close to 0.25 or 0.75.
The phases are marked by numbers in individual panels.
For the three quadruple systems, ET~Boo, VW~LMi and TV~UMi,
we selected the BF's showing the second, non-eclipsing pair.
While the companion binary of VW~LMi
has an almost circular orbit and its lines are separated during most
of the orbit, the lines of the companion binaries of ET~Boo and TV~UMi
can be separated only during short intervals during the
respective periastron passages.
}

\figcaption[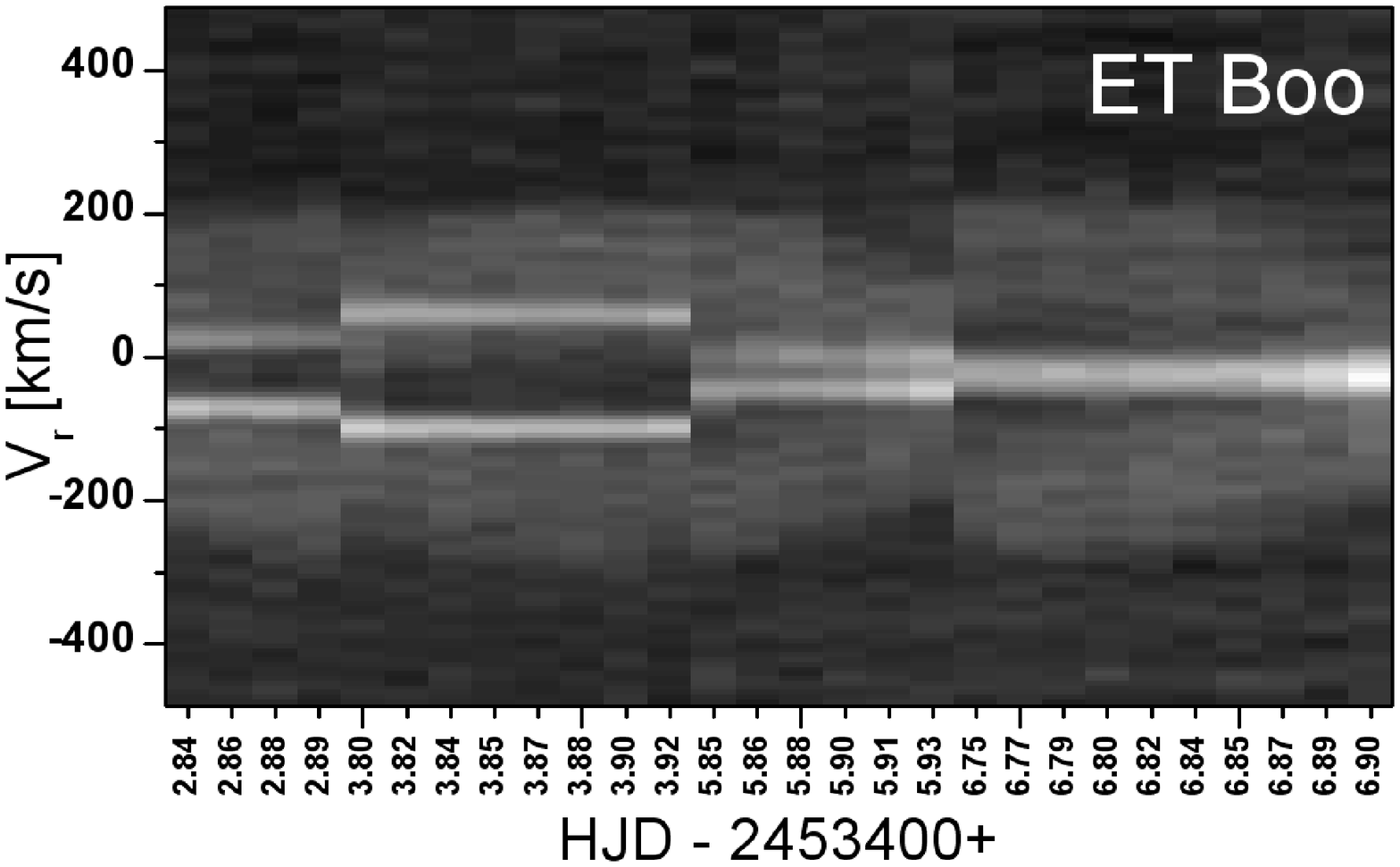] {\label{fig5}
A grayscale plot of broadening functions of ET~Boo extracted from spectra
taken in February, 2005 around 6400~\AA. The spectra were sorted in time.
One bin in X-axis corresponds to one spectrum, so the scale is
not continuous.
The signatures of the close, short-period binary are seen as relatively
faint features in the background. The second binary is visible as a sharp
feature in the BF's which split into two components during the
periastron passage on the second night of observations, on February 10, 2005.
}

\figcaption[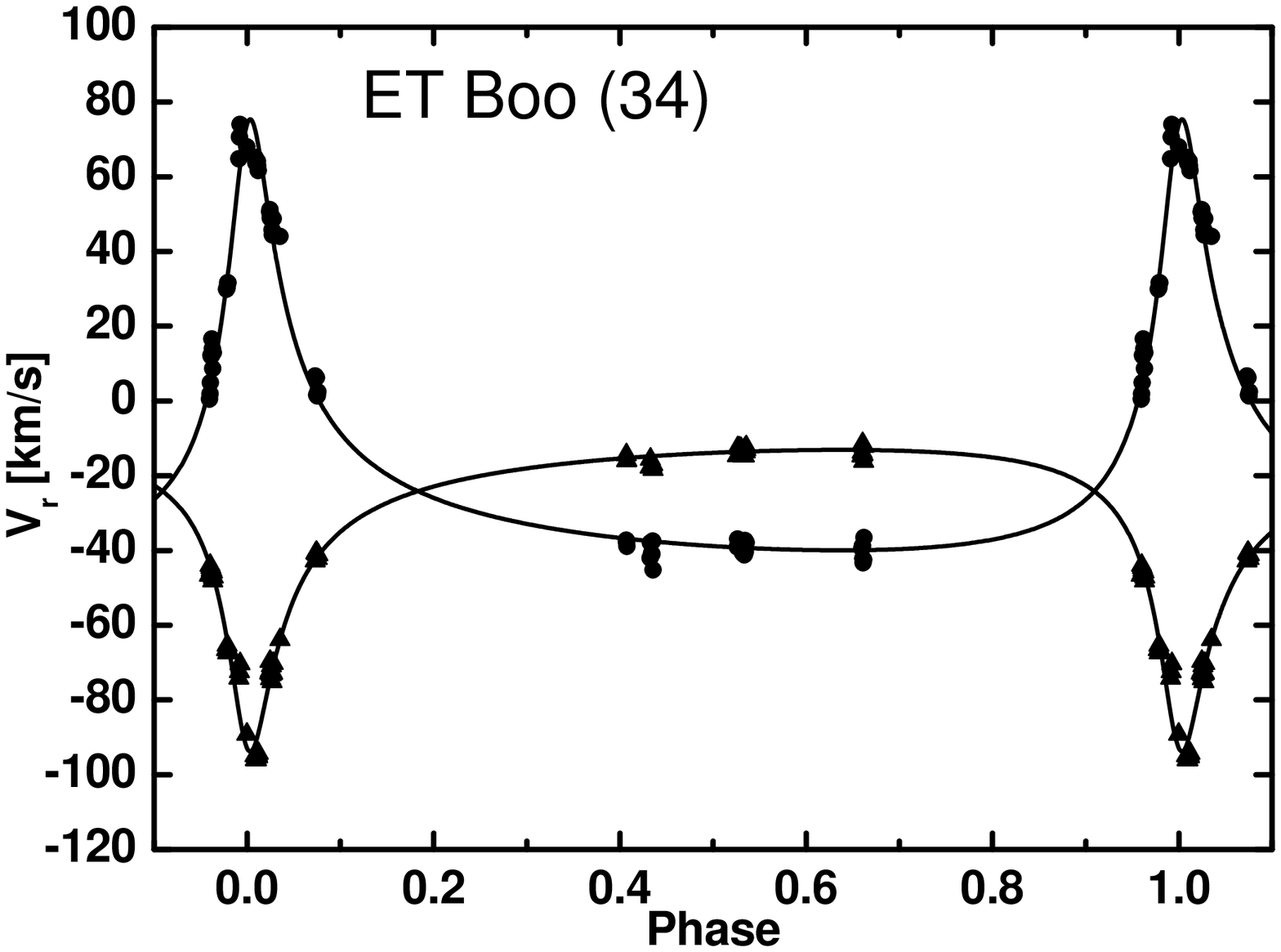] {\label{fig6}
The radial velocities of the third and fourth component of ET~Boo and
the corresponding fits to an orbital motion with the period of 31.52 days.
}

\figcaption[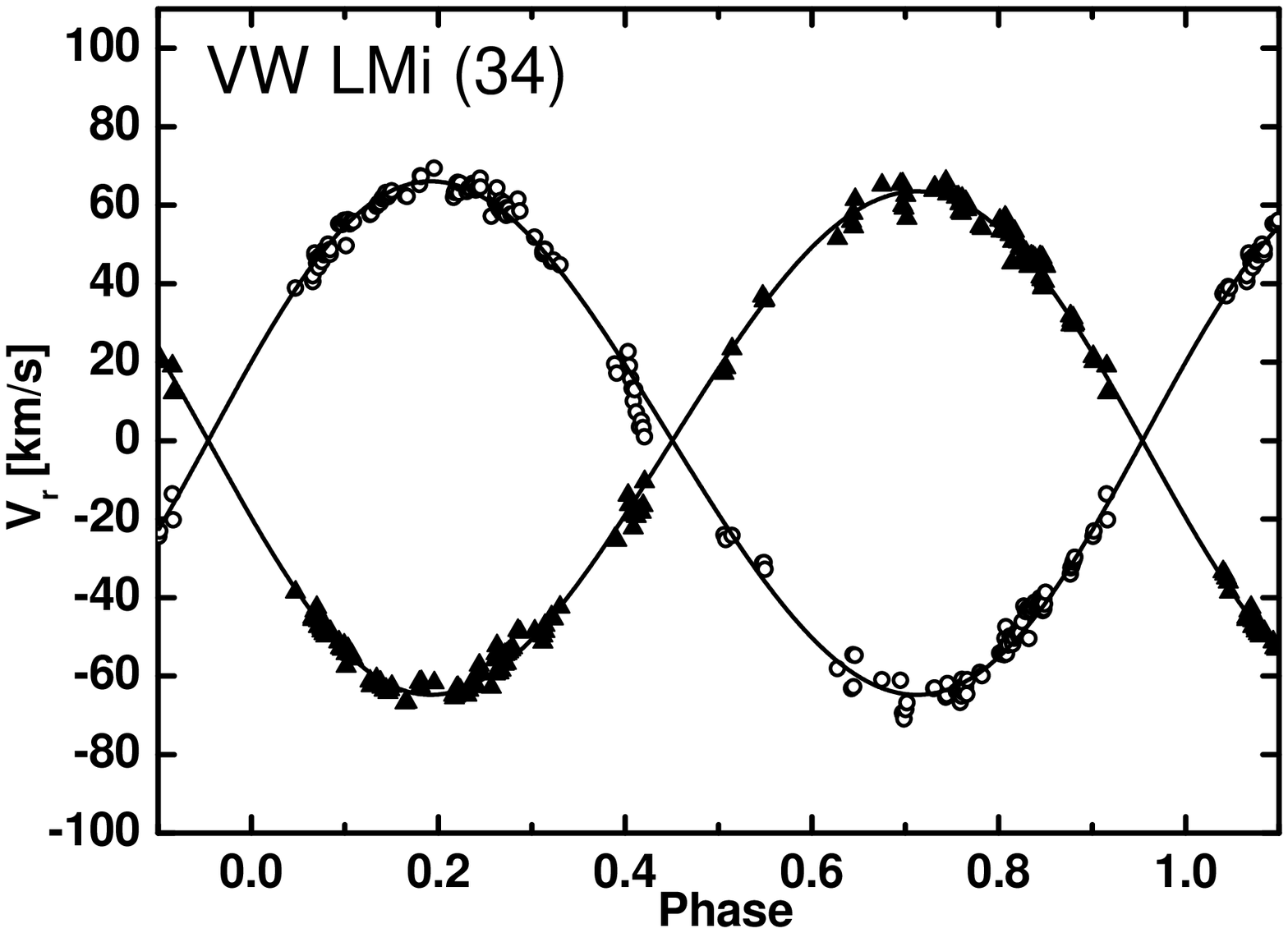] {\label{fig7} The radial velocities of
the third and fourth component of VW~LMi, plotted in a phase
diagram with the period of 7.93 days,  after being corrected for
the motion on the outer, 355 day period orbit.}

\figcaption[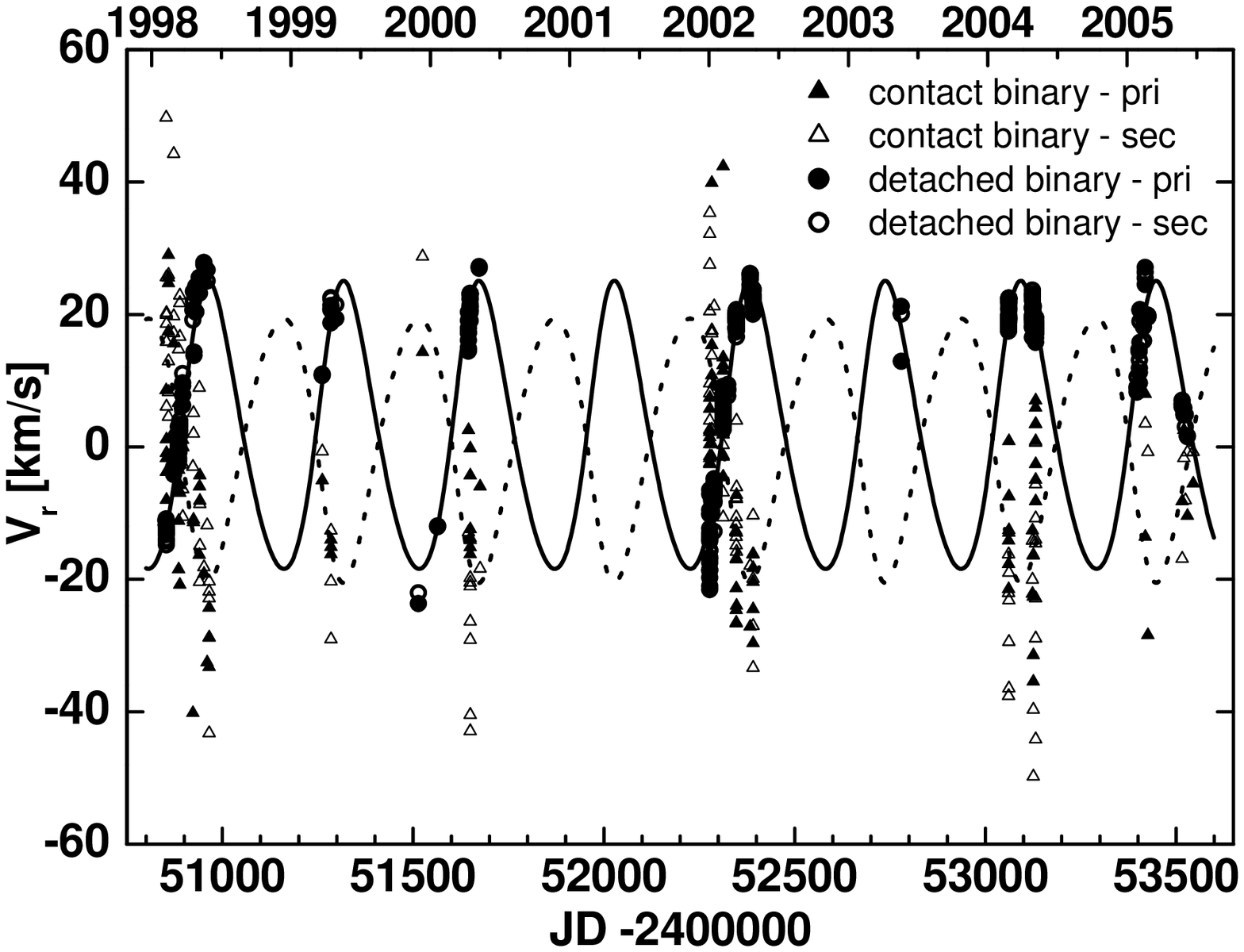] {\label{fig8}
Radial velocities of all four components of VW~LMi corrected for
orbital motion in both close orbits showing motion of the mass centers
in the wide, 355 days orbit. The phases are counted from the periastron
passage. Because the orbital period is close to one year, the
seasonal observing interval slowly moved in the orbital
phases between 1998 and 2005.}

\figcaption[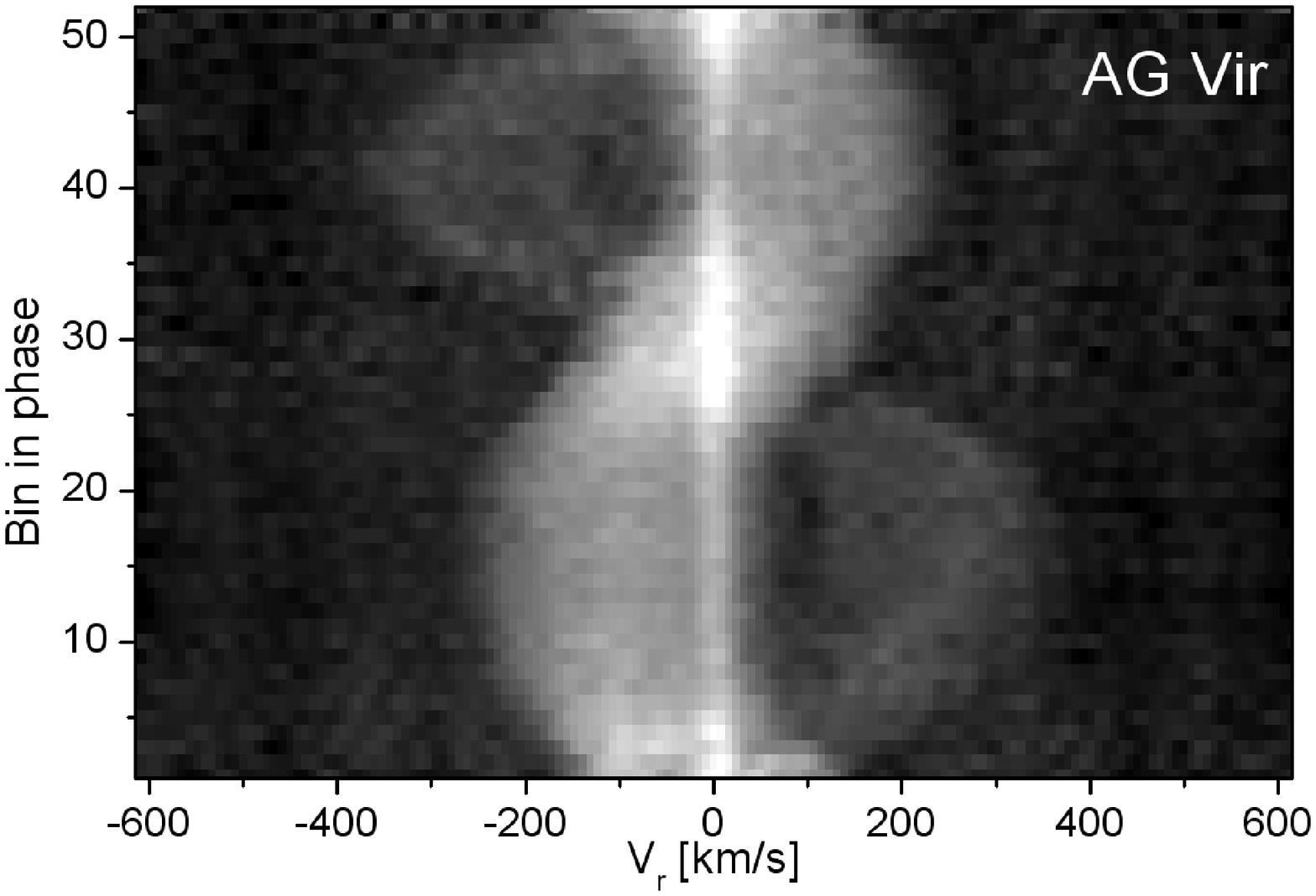] {\label{fig9} A grayscale plot of the
broadening functions of AG~Vir. The spectra have been sorted in
phase, but not rebinned to the proper phase durations. While
individual broadening functions do not show the third component
very clearly, the grayscale plot shows a continuous, bright ridge
line close to the systemic velocity. }

\figcaption[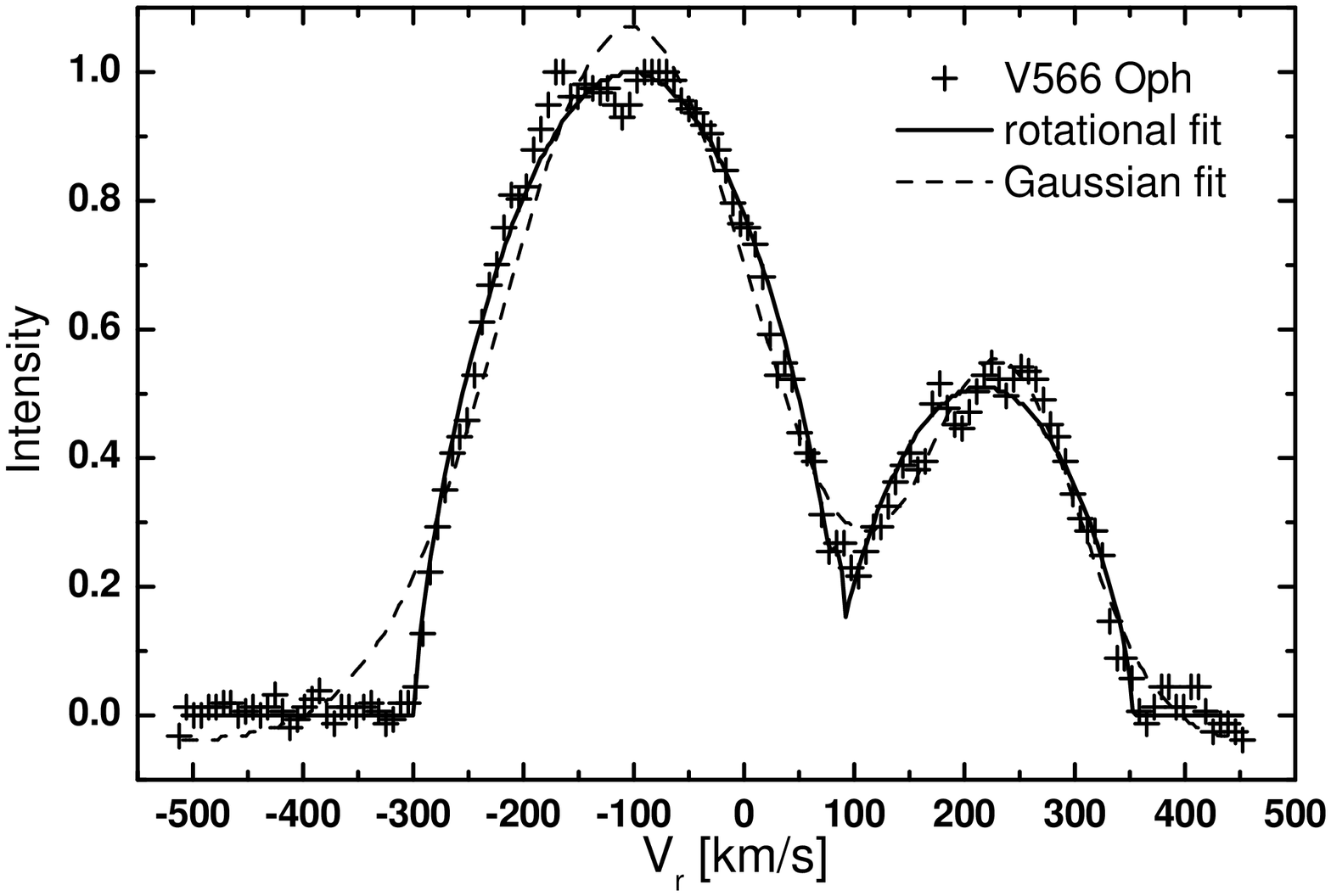] {\label{fig10}
A comparison of double Gaussian and rotational fits to the broadening function
of V566~Oph extracted from spectrum taken close to the orbital
quadrature. The RV is better defined in the steep portions of the
rotational profile branches.
}

\clearpage
\plotone{f1.eps}
% fig.1

\clearpage
\plotone{f2.eps}
% fig.2

\clearpage
\plotone{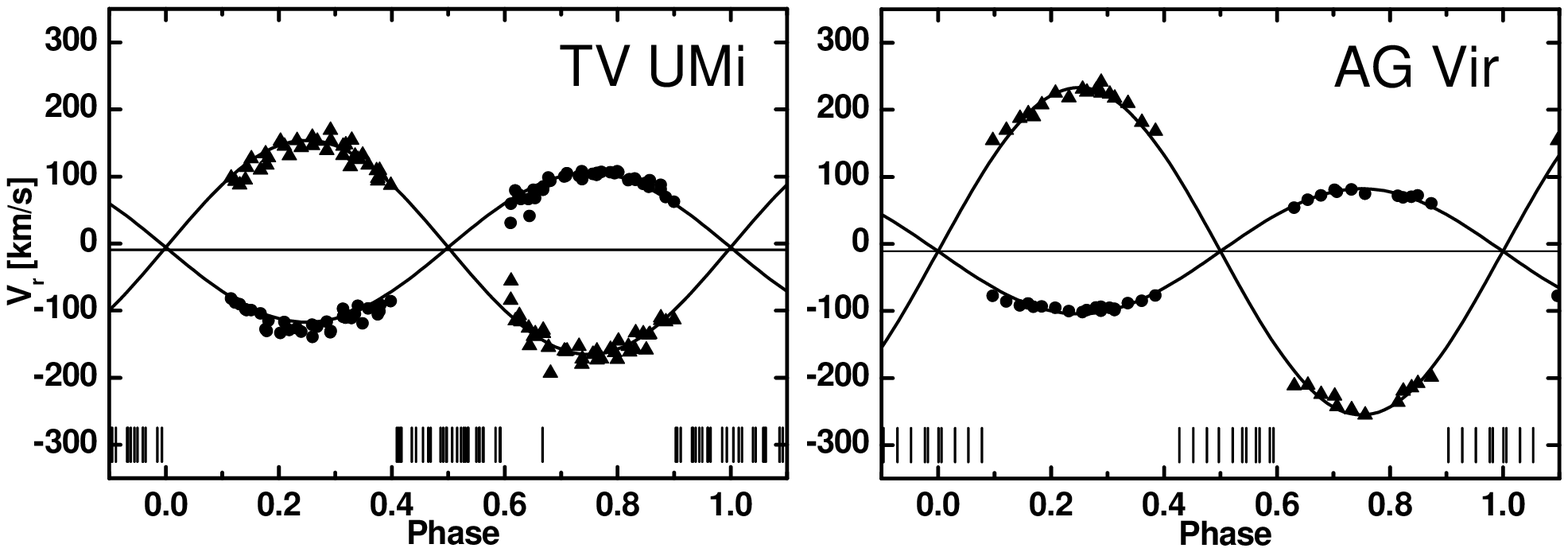}
% fig.3

\clearpage
\plotone{f4.eps}
% fig.4

\clearpage
\plotone{f5.eps}
% fig.5

\clearpage
\plotone{f6.eps}
% fig.6

\clearpage
\plotone{f7.eps}
% fig.7

\clearpage
\plotone{f8.eps}
% fig.8

\clearpage
\plotone{f9.eps}
% fig.9

\clearpage
\plotone{f10.eps}
% fig.10

\clearpage

%----------------------------------------------------------------------
\begin{deluxetable}{lrrrrr}

%\small
\tabletypesize{\footnotesize}
%\tabletypesize{\scriptsize}

\tablewidth{0pt}
%\tablewidth{320pt}
\tablenum{1}

\tablecaption{DDO radial velocity observations of close
binary systems (the full table is available only in electronic form
\label{tab1}}
\tablehead{
\colhead{HJD--2,400,000}  &
\colhead{~V$_1$}          &
\colhead{~~W$_1$}         &
\colhead{~V$_2$}          &
\colhead{~~W$_2$}         &
\colhead{Phase}          \\
                          &
\colhead{[km s$^{-1}$]}   &
                          &
\colhead{[km s$^{-1}$]}   &
                          & \\
}
\startdata
%\sidehead{\bf DU~Boo} \\
 53480.5576  &  22.48~ &  0.35 & $-168.48$~ &  0.03 &  0.6224 \\
 53480.5803  &  25.78~ &  0.47 & $-203.77$~ &  0.03 &  0.6439 \\
 53480.6033  &  31.53~ &  1.49 & $-209.61$~ &  0.14 &  0.6657 \\
 53480.6261  &  34.74~ &  1.09 & $-216.95$~ &  0.05 &  0.6873 \\
 53480.6490  &  33.67~ &  1.05 & $-227.17$~ &  0.08 &  0.7090 \\
 53480.6717  &  40.69~ &  1.00 & $-233.93$~ &  0.30 &  0.7305 \\
 53480.6944  &  37.69~ &  1.19 & $-235.97$~ &  0.13 &  0.7520 \\
 53480.7172  &  38.72~ &  0.89 & $-251.40$~ &  0.13 &  0.7736 \\
 53480.7399  &  38.37~ &  1.03 & $-237.67$~ &  0.10 &  0.7951 \\
 53480.7626  &  40.39~ &  1.32 & $-221.31$~ &  0.12 &  0.8165 \\
\enddata

\tablecomments{The table gives the RVs $V_i$ and
associated weights $W_i$ for observations described in the paper.
The first 10 rows of the table for the first program star, DU~Boo,
are shown. Observations leading to entirely inseparable broadening
function peaks are given zero weight; these observations may be
eventually used in more extensive modelling of broadening
functions. The observations in the clumn $V_1$ correspond to
the component which was stronger and easier to measure in the
analysis of the broadening functions; it was not always the
component eclipsed during the primary minimum at the epoch $T_0$
(see Table~2). The figures should help in identifying which star
is which.}

\end{deluxetable}

%-------------------------------------------------------------------------
\begin{deluxetable}{lccrrrccc}

%\tabletypesize{\small}
%\tabletypesize{\footnotesize}
\tabletypesize{\scriptsize}

\pagestyle{empty}
\tablecolumns{9}

%\tablewidth{660pt}
\tablewidth{0pt}

\tablenum{2}
\tablecaption{Spectroscopic orbital elements \label{tab2}}
\tablehead{
   \colhead{Name} &                % 1
   \colhead{Type} &                % 2
   \colhead{Other names} &         % 3
   \colhead{$V_0$~~~} &            % 4
   \colhead{$K_1$~~~} &            % 5
   \colhead{$\epsilon_1$~} &       % 6
   \colhead{T$_0$ -- 2,400,000} &  % 7
   \colhead{P (days)} &            % 8
   \colhead{$q$}          \\       % 9
   \colhead{}     &                % 1
   \colhead{Sp.~type}    &         % 2
   \colhead{}      &               % 3
   \colhead{} &                    % 4
   \colhead{$K_2$~~~} &            % 5
   \colhead{$\epsilon_2$~} &       % 6
   \colhead{$(O-C)$(d)~[E]} &      % 7
   \colhead{$(M_1+M_2) \sin^3i$} & % 8
   \colhead{}                      % 9
}
% format template
% (1) name, sp & (2) type           & (3) name HD/BD   & (4) V0   & (5) K1 &
% (6) errV1    & (7) T0-2,400,000   & (8) P            & (9) q    \\
% (1)          & (2) sp type        & (3) BD/Hip       &  (4)     & (5) K2 &
% (6) errV2    & (7) O-C ~ [E]      & (8) (M1+M2)sin3i & (9)      \\
\startdata
DU Boo     & EW(A)            & HD~126080  & $-13.09$(0.86) & 53.59(1.02)&
      5.95 & 53,494.6896(39)  & 1.05588870 & 0.234(35) \\ %28 spectra from 2005, 5184A
           & A7V              & HIP~70240  &           & 229.32(3.02)    &
     11.22 & $+0.0068$~[1173] & 2.477(73)  &           \\[1mm]

ET Boo     & EB               & BD+47~2190 & $-23.35$(0.51) &145.61(0.75)&
      9.09 & 52701.5928(8)    & 0.6450398(7)& 0.884(15) \\%158 spectra from several years
           & F7V              & HIP~73346  &           & 164.64(0.93)    &
     10.09 & $-0.0008$~[312]  & 1.996(17)  &           \\[1mm]

TX Cnc     & EW(W)            &            &$+33.97$(0.52) &100.77(0.82) &
     4.09  & 51,807.9810(4)   & 0.38288273 &0.455(11)  \\ % 25 spectra from 2004
           & (F8V)            & BD+19~2068 &           &  221.67(0.91)   &
     5.66  & $+0.0008$[-1807.5]& 1.330(12)  &           \\[1mm]

V1073 Cyg  & EW(A)            & HD~204038  &$-6.85$(0.50)  &66.45(0.61)  &
     4.81  & 53,194.3876(12)  & 0.7858506  &0.303(17)  \\ % 46/45 spectra from 2004
           & A9V              & HIP~105739 &           &  219.09(1.40)   &
     6.63  & $+0.0025$ [883]  & 1.896(25)  &           \\[1mm]

HL Dra     & EB(SB1)          & HD~172022  & $-29.36$(0.43)  & 81.04(0.60)&
      1.88 & 53,166.8057(12)  & 0.944276   &           \\ %33 spectra from 2004 at 5184 A
           & A6V              & HIP~91052  &           &                  &
           & $+0.0047$~[4942] &            &           \\[1mm]

AK Her     & EW(A)            & HD~155937  & $+4.28$(0.89) & 70.52(1.12) &
     6.90  & 53176.3946(19)   & 0.4215231  & 0.277(24) \\ % 38 spectra from 2004
           & F4V              & HIP~84293  &           &  254.40(2.27)   &
    13.90  & $+0.0009$~[1,604]& 1.598(29)  &           \\[1mm]

VW LMi  \tablenotemark{a}     & EW(A)            & HD~95660   &  --~~~~  &105.41(0.83)  &
    12.68  & 51,973.4117(4)   & 0.4775499(2) &0.416(4)  \\
           & F5V              & HIP~54003  &           &  253.21(0.84)   &
    13.58  & $-0.0017$[-1103] & 2.282(18)  &           \\[1mm]

V566 Oph   & EW(A)            & HD~163611  & $-37.33$(0.52) &
71.08(0.69)&
     3.12  & 53,568.6298(4)   & 0.4096538  & 0.263(12) \\ % 33/32 spectra from 2005, 5184 A
           & F4V              & HIP~87860  &           & 270.12(1.14)    &
     6.63  & $+0.0037$~[2608] & 1.686(17)&           \\[1mm]

TV UMi     & EW(W)            & HD~133767  & $-9.70$(0.67) & 116.17(1.04)&
    7.57   & 52,454.0189(6)   & 0.41554935 & 0.739(21) \\ % 74/73 spectra used
           & F8V              & HIP~73474  &           & 157.09(1.19)    &
    9.98   & $+0.0007$~[-111.5]& 0.879(12) &           \\[1mm]

AG Vir     & EW(A)            & HD~104350  & $-10.99$(0.82) & 93.39(1.06)&
     5.38  & 53,501.5388(13)  & 0.6426507  & 0.382(21)  \\  %31 spectra from 2005, 5184 A
           & A5V              & HIP~58605  &           & 244.24(1.97)    &
    10.81  & $+0.0030$~[1558] & 2.563(41)  &           \\[1mm]

\enddata

\tablenotetext{a}{VW~LMi: The contact binary revolves in
relatively short-period orbit around the second pair, therefore
systemic velocity $V_0$ is not defined. See Table~\ref{tab5} for a
full description of the system.}

\tablecomments{The spectral types given in column two
relate to the combined spectral type of all components
in a system; they are given in parentheses if
taken from the literature, otherwise are new,
The convention of naming the
binary components in the table is that the more massive
star is marked by the subscript ``1'', so that
the mass ratio is defined to be always $q \le 1$. The figures
should help identify which component is eclipsed at the primary
minimum. The standard errors of the circular
solutions in the table are expressed in units of last decimal places
quoted; they are given in parentheses after each value.
The center-of-mass velocities ($V_0$), the
velocity amplitudes ($K_i$) and the standard unit-weight
errors of the solutions ($\epsilon$) are all expressed
in km~s$^{-1}$. The spectroscopically determined
moments of primary minima are given by $T_0$; the corresponding
$(O-C)$ deviations (in days) have been calculated from the
available prediction on $T_0$, as given in the text,
using the assumed periods and the number of epochs given by [E].
The values of $(M_1+M_2)\sin^3i$ are in the solar mass units.\\
Ephemerides (HJD$_{min}$ -- 2,400,000 + period in days) used
for the computation of $(O-C)$ residuals:
DU~Boo: 52256.1254 + 1.0558887; ET~Boo: 52500.3407 + 0.6450413;
TX~Cnc: 52500.0407 + 0.38288273; V1073~Cyg: 52500.479 + 0.7858506;
HL~Dra: 48500.189 + 0.944276; AK~Her: 52500.2709 + 0.42152292;
VW~LMi: 52500.1516 + 0.4775505; V566~Oph: 52500.249 + 0.4096538;
TV~UMi: 52500.352 + 0.41555; AG~Vir: 52500.286 + 0.6426507;
}

\end{deluxetable}

%----------------------------------------------------------------------
\begin{deluxetable}{lrrc}

%\small
\tabletypesize{\footnotesize}
%\tabletypesize{\scriptsize}

\tablewidth{0pt}
%\tablewidth{320pt}
\tablenum{3}
\tablecolumns{4}

\tablecaption{Radial velocity observations of the third and fourth
components of quadruple systems (the full
table is available only in electronic form) \label{tab3}}
\tablehead{
\colhead{HJD--2,400,000} & \colhead{~V$_3$}         & \colhead{~V$_4$}        & \\
                         & \colhead{[km s$^{-1}$]}  & \colhead{[km s$^{-1}$]} & \\
}
\startdata
\sidehead{\bf ET~Boo}
53455.89155 & $-14.30$~~ &  $-42.26$~~ & b \\
53455.90414 & $-14.30$~~ &  $-43.31$~~ & b \\
53455.91510 & $-16.13$~~ &  $-42.56$~~ & b \\
53455.92738 & $-12.09$~~ &  $-36.59$~~ & b \\
53529.61628 & $-89.18$~~ &  $ 68.01$~~ &   \\
53530.73133 & $-63.86$~~ &  $ 44.00$~~ &   \\
\sidehead{\bf VW~LMi}
50852.78735 & $-75.68$~~ &  $ 51.31$~~ &   \\
50852.79458 & $-75.45$~~ &  $ 52.17$~~ &   \\
50853.69796 & $-54.16$~~ &  $ 29.50$~~ &   \\
50853.70383 & $-55.28$~~ &  $ 28.85$~~ &   \\
\enddata

\tablecomments{The table gives the RVs $V_i$ for
the third and fourth components. The typical 10 rows of the table
for the quadruple systems, ET~Boo and VW~LMi, are shown. Observations of
quadruple systems ET~Boo, VW~LMi and TV~UMi leading to entirely
inseparable broadening function peaks of components of the second
binary are omitted from the table. Radial velocities determined from
partially blended profiles are marked by ``b'' in the last column.}
\end{deluxetable}

%----------------------------------------------------------------------
\begin{deluxetable}{lrr}

%\small
\tabletypesize{\footnotesize}
%\tabletypesize{\scriptsize}

\tablewidth{0pt}
%\tablewidth{320pt}
\tablenum{4}

\tablecaption{Spectroscopic orbital elements of the second
non-eclipsing binary in the quadruple system ET~Boo \label{tab4}}
\tablehead{
\colhead{Parameter} &  & \colhead{error} \\
}
\startdata
$P_{34}$ [days]        & 31.5212  &   0.0005   \\
$e_{34}$               & 0.740    &   0.011    \\
$\omega$ [rad]         & 2.94     &   0.03     \\
$T_0$ [HJD]            & 2,451,354.66 & 0.04 \\
$V_0$ [km~s$^{-1}$]     &   $-$24.15 &   0.44    \\
$K_3$ [km~s$^{-1}$]     &    40.36 &   0.67    \\
$K_4$ [km~s$^{-1}$]     &    57.67 &   0.69    \\
$q = K_4/K_3$          &    0.70  &   0.09     \\
$(a_3+a_4)\sin i$ [AU] &    0.191 &   0.003    \\
$(M_3+M_4)\sin^3 i$ [M$_\odot$] & 0.93 & 0.05  \\
\enddata

\tablecomments{The table gives spectroscopic elements of the second
 binary in ET~Boo: orbital period ($P_34$), eccentricity ($e_{34}$),
 longitude of the periastron passage ($\omega$), time of the periastron
 passage ($T_0$), systemic velocity ($V_0$), semi-amplitudes of the
 RV changes ($K_3,K_4$). Corresponding mass ratio
 $q$, and projected relative semi-major ($(a_3+a_4)\sin i$) and
 total mass ($(M_3+M_4)\sin^3 i$) is also given.}

\end{deluxetable}

%----------------------------------------------------------------------
\begin{deluxetable}{lrr}

%\small
\tabletypesize{\footnotesize}
%\tabletypesize{\scriptsize}

\tablewidth{0pt}
%\tablewidth{320pt}
\tablenum{5}

\tablecaption{Spectroscopic orbital elements of all three orbits
              defined by RVs of four resolved components of
              VW~LMi \label{tab5}}
\tablehead{
\colhead{Parameter} &  & \colhead{error} \\
}
\startdata
\multicolumn{3}{l}{\bf Contact (eclipsing) pair - circular orbit} \\
$P_{12}$ [days]           & 0.47754988 &   0.00000020 \\
$T_0$ [HJD]          & 2,451,973.4117 & 0.0004 \\
$K_1$ [km~s$^{-1}$]  &   105.41   &   0.83       \\
$K_2$ [km~s$^{-1}$]  &   253.21   &   0.84       \\
$(M_1+M_2)\sin^3 i_{12}$ [M$_\odot$] &  2.282  &  0.018  \\
\hline
\multicolumn{3}{l}{\bf Detached non-eclipsing pair}    \\
$P_{34}$ [days]          & 7.93062    &   0.00014    \\
$e_2$                & 0.033      &   0.010      \\
$\omega_2$ [rad]     & 5.01       &   0.09       \\
$T_0$ [HJD]          & 2,452,282.44 & 0.11     \\
$K_3$ [km~s$^{-1}$]  & 65.44      &   0.76       \\
$K_4$ [km~s$^{-1}$]  & 64.15      &   0.76       \\
$(M_3+M_4)\sin^3 i_{34}$ [M$_\odot$] &  1.788  &  0.033  \\
\hline
\multicolumn{3}{l}{\bf Mutual wide orbit}              \\
$P_{12-34}$ [days]      & 355.0      &   0.5        \\
$e_3$             &   0.14     &   0.03       \\
$\omega_3 $ [rad] &   2.35     &   0.22       \\
$T_0$ [HJD]            & 2,452,703&  11          \\
$K_{12}$ [km~s$^{-1}$] & 19.96      &   0.84       \\
$K_{34}$ [km~s$^{-1}$] & 21.76      &   0.79       \\
$(M_1+M_2+M_3+M_4)\sin^3 i_{12-34}$ [M$_\odot$] &  2.67  &  0.16 \\
\hline
$V_0$ [km~s$^{-1}$]    &  1.29      &   0.39       \\
\enddata

\tablecomments{The table gives spectroscopic elements for the
three observed orbits of VW~LMi.
The designation of parameters is as in the previous table.
The index ``12'' refers to the orbit of the contact pair, while the
index ``34'' refers to the orbit of the second, detached binary.
parameters of the mutual orbit of
these binaries are indexed as ``12-34''.
The orbit of the contact pair is assumed
to be circular ($e_1$=0, $\omega_1 = \pi/2$).}
\end{deluxetable}

\end{document}